\documentclass{main}
\usepackage[most]{tcolorbox}
\usepackage{slashed}
\usepackage{amsmath}
\usepackage{textcomp}
\usepackage{fancyhdr}
\usepackage{lastpage}
\usepackage{siunitx}
\usepackage{soul}
\usepackage{cite}
\usepackage[dvipsnames]{xcolor}

\def\be{\begin{equation}}
\def\ee{\end{equation}}
\def\bea{\begin{eqnarray}}
\def\eea{\end{eqnarray}}

\fancypagestyle{firstpagefooter}{%
  \fancyfoot[L]{  \textcopyright \footnotesize This work is an open access article under a Creative Commons Attribution 4.0 International License (https://creativecommons.org/licenses/by/4.0/).}
  \fancyfoot[C]{}  
   
}

\newcommand{\FIG}{Fig.~}

\newcommand{\EQ}{Eq.~}
\newcommand{\EQs}{Eqs.~}

\begin{document}

\thispagestyle{firstpagefooter}
\title{\Large EXCLUSIVE PHOTOPRODUCTION OF A DI-MESON PAIR WITH LARGE INVARIANT MASS }

\author{\underline{DAVID PEREZ}\footnote{Speaker, email: david.perez@ijclab.in2p3.fr}}

\address{
Université Paris-Saclay, CNRS/IN2P3, IJCLab, Orsay, 91405, France
}
\author{\underline{SAAD NABEEBACCUS}}
\address{Department of Physics and Astronomy, University of Manchester, Manchester M13 9PL, United Kingdom}

\author{\underline{LECH SZYMANOWSKI}}
\address{National Center of Nuclear Research (NCBJ), Pasteura 7,Warsaw, 02-093, Poland}

\author{\underline{SAMUEL WALLON}}
\address{Université Paris-Saclay, CNRS/IN2P3, IJCLab, Orsay, 91405, France}

\maketitle\abstracts{We consider the exclusive photoproduction of a di-meson pair with large invariant mass, $\gamma N \rightarrow N' M_1M_2$, in the framework of collinear factorisation. The mesons considered $M_1$ and $M_2$ are either pions or rho mesons, charged or neutral.
We consider the kinematic regime characterised by a large invariant mass of the two-meson system, and a small deflection of the nucleon in the centre-of-mass frame. In this kinematic domain, the amplitude factorises into a perturbative hard part and non-perturbative parts described by Generalised Parton Distributions (GPDs) and Distribution Amplitudes (DAs). We automate the calculation of the fully differential cross section at leading twist and leading order, and we present some numerical results at JLab 12 GeV kinematics. This class of processes provides yet more exclusive $2 \to 3$ channels that can be used to extract GPDs.}

\setlength{\parskip}{2mm}


\section{Introduction}
Understanding the internal dynamics of nucleons has been one of the most active research fields for the last few decades. Beyond the Parton Distribution Functions (PDFs) which describe the probability of finding a parton carrying longitudinal momentum fraction $x$ inside the nucleon, it has now become possible to access the transverse spatial distribution  through Generalised Parton Distributions (GPDs) \cite{Burkardt:2000za,Burkardt:2002hr}.  These functions appear as the non-perturbative part of the amplitude for exclusive processes where collinear factorisation is expected to hold.

Here, we focus our attention on the family of $2 \to 3$ exclusive processes of the form $\gamma N\rightarrow N' M_1 M_2$, 
namely the exclusive photoproduction of a di-meson pair \cite{di-meson-paper}. 
We work in the kinematics where the di-meson pair has a large invariant mass (provided by the large relative transverse momentum of the final-state mesons), and small squared momentum transfer to the nucleon. In such a case, it can be shown that the process admits a collinear factorisation approach, and thus provides access to GPDs \cite{Qiu:2022bpq,Qiu:2022pla}. Depending on the spin and polarisation of the two mesons, chiral-even (helicity non-flip) and/or chiral-odd (helicity-flip) GPDs contribute to the amplitude.

Our work aims to provide theoretical predictions for the considered class of exclusive processes, which will be indispensable for future experimental studies, by assessing their sensitivity to GPD modeling, especially with respect to chiral-odd GPD which are poorly constrained experimentally. In this report, we restrict ourselves to Leading-Order (LO) in $\alpha_s$ at the leading twist, and to processes that do not involve 2-gluon exchanges in the $t$-channel, in order to avoid the factorisation breaking effects described in \cite{Nabeebaccus:2023rzr,Nabeebaccus:2024mia}.

\section{Kinematics}

We label the momenta and polarisation vectors in the process as follows:
\begin{equation}
    \gamma(q,\epsilon_{q})+N(p_1)\rightarrow N'(p_2)+M_1(p_{M_1},\epsilon_{M_1})+M_2(p_{M_2},\epsilon_{M_2})\,,
\end{equation}
where $\epsilon_1$ and $\epsilon_2$ are the polarisations of the two outgoing mesons if they are vector mesons.

We define the momenta
\begin{equation}
    q^\mu=\frac{\sqrt{s}}{2}(1,0,0,-1)\,,\qquad
   p^\mu=\frac{\sqrt{s}}{2} (1,0,0,1)\,.
\end{equation}
 Any vector $v^\mu$ can be decomposed in the Sudakov basis as
 \begin{equation}
     v^\mu=v^+p^\mu +v^-q^\mu+v_{\perp}^\mu\,.
 \end{equation}
We further define $\Delta=p_2-p_1$, and  introduce the usual skewness parameter $\xi=-\frac{\Delta^+}{2 p^+}$.

In the limit where the deflection of the nucleon is negligible ($\Delta_{\perp}$ small compared to the hard scale in the process), and if we neglect all hadron masses, the momenta of the particles in the process are given by
\begin{equation}
    p_1^{\mu}=(1+\xi)\,p^{\mu},\quad\; p_2^{\mu}=(1-\xi)\,p^{\mu}\,,\quad\;
      p_{M_1}^{\mu}=(1-\alpha)\, q^{\mu}+2\alpha \xi \,p^{\mu}-p^{\mu}_\perp\,,\quad\;
    p_{M_2}^{\mu}=\alpha\, q^{\mu} + 2(1-\alpha)\xi \,p^{\mu}+p^{\mu}_\perp\,.
\end{equation}
One can show (see \cite{Duplancic:2023kwe} for details) that in this limit, we have:
\begin{align}
    &S_{\gamma N}\equiv(q+p_1)^2=(1+\xi)s\,,\\
    &M_{12}^2\equiv(p_{M_1}+p_{M_2})^2=2\xi s\,.
\end{align}
From these equations, it follows that
\begin{align}
\label{eq:t-primed}
    -t'&\equiv-(q-p_{M_2})^2=(1-\alpha)\,M_{12}^2\,,\\
    \label{eq:u-primed}
    -u'&\equiv -(q-p_{M_1})^2=\alpha\,M_{12}^2\,,\\
    s' &\equiv (p_{M_1}+p_{M_2})^2 = M_{12}^2=\frac{\Vec{p_t}^2}{\alpha(1-\alpha)}\,.
    \label{eq:invariant-mass}
\end{align}
Since $M_{12}^2$ is positive,  the last equality implies that $0<\alpha<1$, so that $-t'$ and $-u'$ are also positive.

The polarisation vectors of the outgoing mesons can be written as \cite{Duplancic:2023kwe}
\begin{align}
\epsilon^{\mu}_{M_{i}}(p_{M_{i}},L) &=\frac{1}{m_{M_{i}}}p^\mu_{M_{i}}-\frac{m_{M_{i}}}{p \cdot p_{M_{i}}}p^\mu\,,\\
    \epsilon^{\mu}_{M_{i}}(p_{M_{i}},T) &= \epsilon_{M_{i}\perp}^{\mu} - \frac{\epsilon_{M_{i}\perp}\cdot p_{M_{i}}}{p \cdot p_{M_{i}}}p^{\mu}\,,
\end{align}
where $L$ and $T$ stand for longitudinal and transverse respectively, and $m_{M_{i}}$ represents the mass of the light meson $M_{i}$.


We choose the axial gauge, $\epsilon_{q}\cdot p = 0 $ for the photon polarisation vector, which leads to
\begin{equation}
    \epsilon_q^{\mu}=\epsilon_{q\perp}^{\mu}\,.
\end{equation}

In the chosen kinematics, the amplitudes for our class of processes are functions of the three parameters $\alpha$, $\xi$, $s$ and of the possible tensor structures (scalar products or Levi-Civita tensors) that can be generated with $p_\perp$, $\epsilon_{1,2\perp}$, $\epsilon_{q\perp}$, $p$ and $q$.

\section{Non-perturbative ingredients}
In order to make predictions, we need to model the non-perturbative parts of the amplitude. 
Within the collinear factorisation framework, the amplitude factorises into a perturbative hard part, denoted by $T_H(x,v,z)$, and a non-perturbative part, which consists of a Generalised Parton Distribution (GPD), $H(x)$, and two Distribution Amplitudes (DAs), $\phi_1(v)$ and $\phi_2(z)$. The GPD encodes the internal dynamics of the nucleon, while the DAs describe those of each meson. In the GPD, $x$ represents the average longitudinal momentum of the parton probed from the nucleon, while in the DAs, $v$ and $z$ represent the momentum fraction of each meson carried by the quark.

The factorised amplitude is then given by
\begin{equation}\label{amplitude}
    i\mathcal{M}=\int_{-1}^{1}dx \int_{0}^1dz\int_0^{1}dv\;T_H(x,v,z)H(x)\phi_1(v)\phi_2(z)\,.
\end{equation}

There are two kinds of quark GPDs, namely chiral-even and chiral-odd GPDs. The first ones describe processes in which the total transfer of helicity between the initial and final nucleon is zero. They can be further distinguished into either vector, denoted by $H^q, E^q$, or axial $\Tilde{H}^q,\Tilde{E^q}$, with $q$ being the flavour of the active parton. In our study, only $H^q$ and $\Tilde{H}^q$ are kept, the other being suppressed  due to multiplication by powers of the skewness $\xi$ in the cross section \cite{Duplancic:2022ffo}. The chiral odd GPDs, denoted by $H_T^q,\,\Tilde{H}_T^q,\,E_T^q$ and $\Tilde{E}_T^q$ describe a transfer of one unit of helicity in the $t$-channel. The dominant contribution comes from the transversity GPD $H_T^q$, the other ones being suppressed by powers of $\xi$.

The GPD and the DAs appear in the amplitude as matrix elements of bilocal operators at light-like separation. For example, for the vector GPD, we have
\begin{equation}
\langle N'(p_2)| \bar{q}\left(-\frac{z}{2}\right)\gamma^+q\left(\frac{z}{2}\right)|N(p_1)\rangle \Big{|}_{\substack{z_\perp=0\\z^+=0}}= \int_{-1}^{1}dx\,e^{-i xP^+ z^-}\bar{u}(p_2) \gamma^+u(p_1)\,H^q(x,\xi,t)\,,\end{equation}
while, for an axial DA, as the one of a $\pi^+$,
\begin{equation}
    \langle \pi^+(p_{\pi})|\bar{u}(y)\gamma^5\gamma^\mu d(-y)|0\rangle=if_\pi p_\pi^\mu\int_0^1dz\,e^{-i(2 z-1)p_\pi\cdot y}\phi_\pi(z)\,,
\end{equation}
where $f_{\pi}$ is the pion decay constant.

Various models for GPDs and DAs exist.
For concreteness, we adopt Radyushkin's Double Distribution Ansatz for the GPDs \cite{Radyushkin:1998es},
\begin{equation}
    {H}^q(x,\xi,t)=\int_{\{|\beta|+|\alpha|\leq 1\}}d\beta\,d\alpha\,\delta(\beta+\xi\alpha-x)F^q(\beta,\alpha,t)\,,
\end{equation}
where $F^q$ is constructed using a profile function $\Pi(\beta,\alpha)=\frac{3}{4}\frac{(1-\beta)^2-\alpha^2}{(1-\beta)^3}$ and PDFs through
\begin{align}
    &F^q(\beta,\alpha,t)=\left(\Pi(\beta,\alpha) {q}(\beta)\Theta(\beta)-\Pi(-\beta,\alpha) {\bar{q}}(-\beta)\Theta(-\beta)\right)\frac{C^2}{(t-C)^2}\,,\\
&\tilde{F}^q(\beta,\alpha,t)=\left(\Pi(\beta,\alpha) {\Delta q}(\beta)\Theta(\beta)-\Pi(-\beta,\alpha) {\Delta\bar{q}}(-\beta)\Theta(-\beta)\right)\frac{C^2}{(t-C)^2}\,,\\
    &F^q_ T(\beta,\alpha,t)=\left(\Pi(\beta,\alpha) {\delta q}(\beta)\Theta(\beta)-\Pi(-\beta,\alpha) {\delta \bar{q}}(-\beta)\Theta(-\beta)\right)\frac{C^2}{(t-C)^2}\,.
\end{align}
In the above, $q$, $\Delta q$ and $\delta q$ denote respectively the unpolarised, polarised and transversity PDF of flavour $q$. The constant $C$ is fixed to $C=\SI{0.71}{GeV^2}$. The parametrisation adopted here is the same as the one used in the study of the photoproduction of a photon-meson pair reported in \cite{Boussarie:2016qop,Duplancic:2018bum,Duplancic:2022ffo,Duplancic:2023kwe}. The manipulation of PDFs is facilitated by the use of the \textit{Mathematica} package \textit{ManeParse} \cite{ManeParse}.

 For the DAs, we use the asymptotic form, given by
\begin{align}
    &\phi_i (v)=6\,v\,(1-v)\,,\qquad i=1,2\,.
\end{align}

\section{Automation of amplitude computation}

\subsection{Generation and storage of the amplitudes}
\begin{figure}[h]
    \centering
    \includegraphics[width=0.9\linewidth]{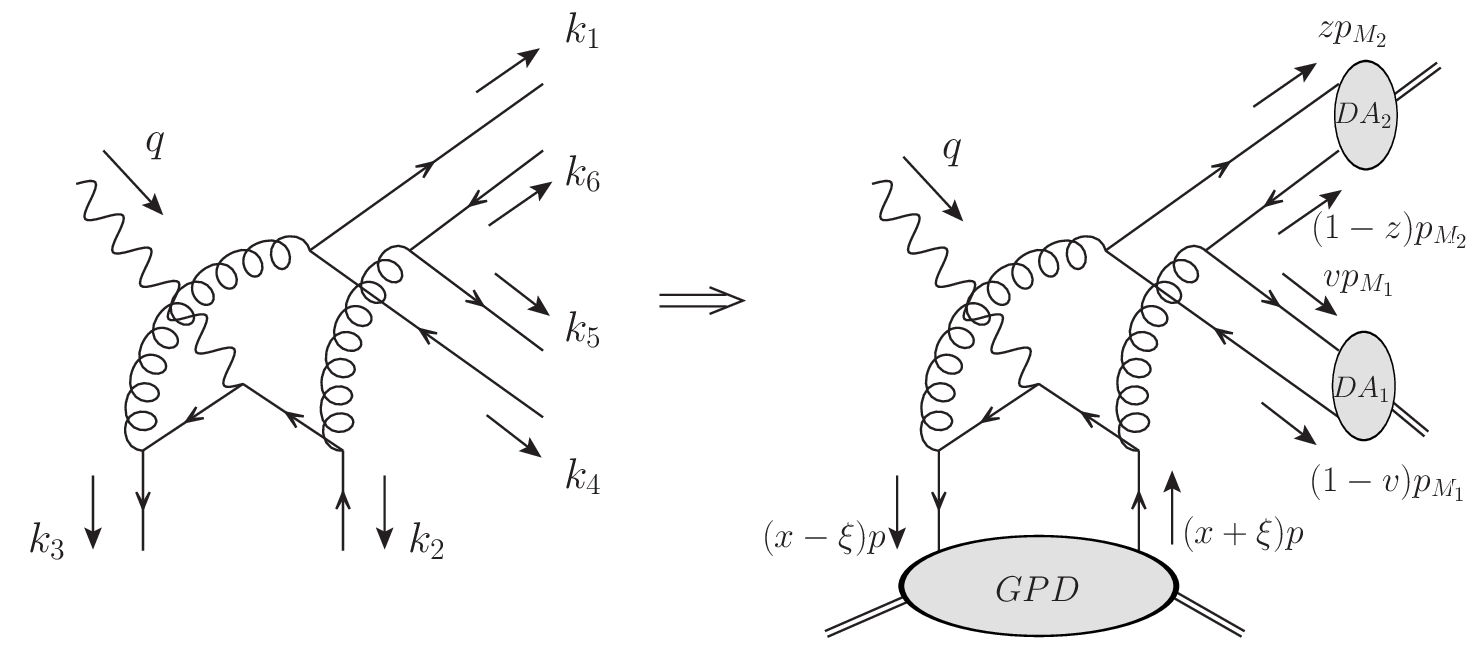}
    \caption{The projection of a diagram obtained with \textit{FeynArts} and \textit{FeynCalc} onto the GPD and DAs.}
    \label{fig:diagramgeneration}
\end{figure}

In order to generate  all possible diagrams without fixing the particle species, we generate the diagrams of the process $\gamma(q)\rightarrow u(k_1)\,\bar{u}(k_2)\,u(k_3)\,\bar{u}(k_4)\,u(k_5)\,\bar{u}(k_6)$  using the \textit{FeynArts} and \textit{FeynCalc} packages \cite{Hahn:2000kx,Shtabovenko:2016sxi,Shtabovenko:2020gxv,Shtabovenko:2023idz,Mertig:1990an} in \textit{Mathematica}. The momenta of the outgoing quarks and anti-quarks are represented by $k_i$, $i=1,...,6$. These diagrams, such as the one shown on the left-hand side of Fig.~\ref{fig:diagramgeneration}, are then mapped onto the desired form by extracting the chains of Dirac matrices between spinors, and assembling them into a Dirac trace. For example, for the specific diagram in \FIG\ref{fig:diagramgeneration},

\begin{equation}
\frac{{\color{black}\bar{u}(k_1)}\pmb{\color{ProcessBlue}{\gamma}_{\mu}}{\color{black}{v(k_4)\bar{u} (k_5)}} \pmb{\color{red}\gamma_\nu}{\color{black} v (k_6)\bar{u} (k_3)}{\pmb{\color{ForestGreen}{\gamma}^\mu(\slashed{k}_1+\slashed{k}_3+\slashed{k}_4)\slashed{\varepsilon}(q1)(-\slashed{k}_2-\slashed{k}_5-\slashed{k}_6)\gamma^\nu}}{\color{black} v (k_2)}}{(k_1+k_4)^2(k_5+k_6)^2(k_2+k_5+k_6)^2(k_1+k_3+k_4)^2}
  \end{equation}
\raisebox{-.5cm}{\makebox[\linewidth][l]{\hspace{0.5\linewidth}\scalebox{1.8}{$\Downarrow$}}}
\begin{equation}
\label{hard-part}
\frac{\mathrm{Tr}\left(\pmb{\color{ForestGreen}\gamma^\mu(\slashed{k}_1+\slashed{k}_3+\slashed{k}_4)\slashed{\varepsilon}(q_1)(-\slashed{k}_2-\slashed{k}_5-\slashed{k}_6)\gamma^\nu }{\color{brown}\boxed{\color{black}GPD}}\right)\mathrm{Tr}\left(\pmb{\color{ProcessBlue} \gamma_\mu} {\color{brown}\boxed{\color{black}DA_1}}\pmb{\color{red}\gamma_\nu}{\color{brown}\boxed{\color{black}DA_2}}\right)}{ (k_1+k_4)^2(k_5+k_6)^2(k_2+k_5+k_6)^2(k_1+k_3+k_4)^2}
\end{equation}

\begin{figure}[ht!]
    \centering
    \includegraphics[width=0.99\linewidth]{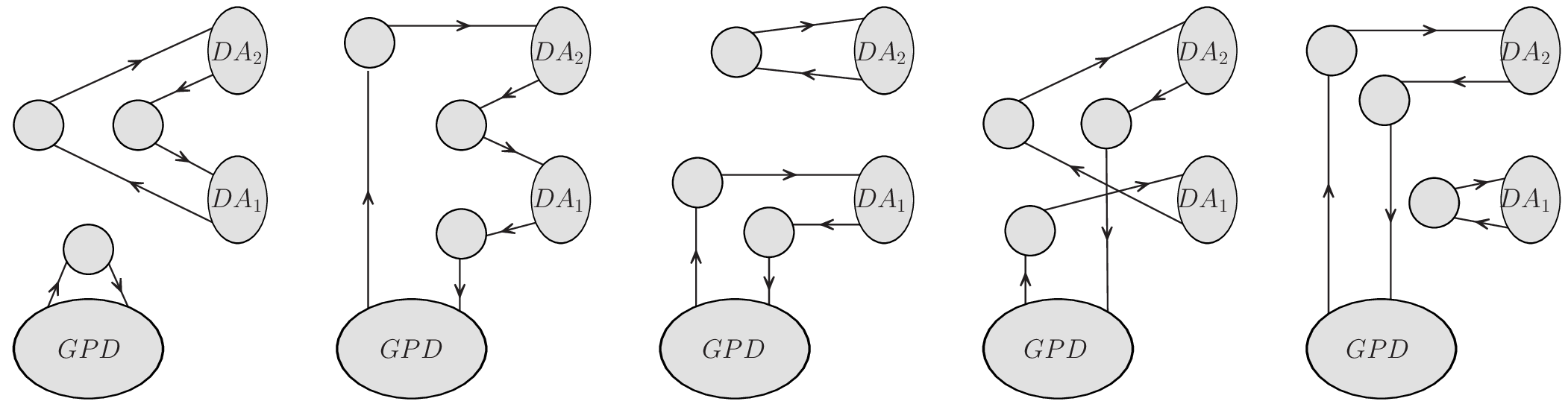}
    \caption{Five ways of connecting the fermion lines, giving five topologies. The unmarked blobs correspond to places where the internal gluons and/or incoming photon could attach to.}
    \label{fig:topologies}
\end{figure}

Each GPD and DA enters the scattering amplitude multiplied by a Dirac matrix, which is absorbed into the hard part $T_H$ in \EQ(\ref{amplitude}). These Dirac matrices are symbolically written as brown boxes in Eq.~(\ref{hard-part}).
Depending on the nature of the GPD or DA, those matrices can be  $\gamma^\mu$ (vector), $\gamma^5\gamma^\mu$ (axial), or $\frac{i}{2}[\gamma^\mu,\gamma^\nu]$ (tensor, which corresponds to transversity). There are $3^3=27$ possible combinations of Dirac structures. However, the tensor Dirac structures must appear in pairs, otherwise the amplitude vanishes due to the presence of an odd number of gamma matrices in the trace. This reduces the number of combinations to 14. Therefore, each diagram on the right-hand side of Fig.~\ref{fig:diagramgeneration} can be projected onto 14 different possible combinations of Dirac structures. This gives a total of 1480 diagrams which are organised according to their topology. The five possible topologies are represented in Fig.~\ref{fig:topologies}.

A flag containing all the structural information is associated with each diagram. As an illustration, a generic amplitude, corresponding to the right panel of \FIG\ref{fig:diagramgeneration}, is given by
\begin{align} \label{example-amplitude}
&\frac{e g^4 C_F
   \left(\epsilon_{q\perp}\cdot p_{\perp}\right)
   \left(\alpha  \xi  \left(-4 v^2+4 v-4 z^2+4 z-2\right)+\xi  (1-2
   v)^2+2 (\alpha -1) v x-2 \alpha  x z+x\right)
   }{64 \xi ^2 {N_c}^2 s^2 (v-1) v
   (z-1) z (\alpha  (v+z-1)-v) (\alpha  (v+z-1)-v+1)
   \left(i\epsilon+\frac{\xi  (-\alpha  (v+z-1)-2 v z+v+2 z-1)}{\alpha 
   (v+z-1)-v+1}+x\right)} \nonumber\\[5pt]
  &\hspace{6cm} \times\frac{f(\{244,1,1,\{3,2\}\},N,N',\{M_1,P_1\},\{M_2,P_2\})}{\left(-i\epsilon+\frac{\xi  (\alpha  (v+z-1)-2
   v z+v)}{\alpha  (v+z-1)-v}+x\right)}\,.
\end{align}
The flag $f(\{244,1,1,\{3,2\}\},N,N',\{M_1,P_1\},\{M_2,P_2\})$ has a list of four arguments as its first entry, which respectively indicates that this amplitude
\begin{itemize}
    \item originates from diagram number 244,
    \item was  calculated using only vector GPD and DAs (combination number 1, out of 14),
    \item  belongs to the first topology  (out of 5) shown in \FIG\ref{fig:topologies}, and
    \item the photon is attached to the fermion line connecting $k_2$ to $k_3$.
\end{itemize}
The variables $N$, $N'$, $M_1$, $M_2$, $P_1$, $P_2$ refer respectively to the incoming and outgoing nucleon species, the first and second type of meson, and their polarisations $P_1$ and $P_2$. Once these are specified, the flag becomes the prefactor of the diagram for the given process. This prefactor includes the electric charge (of the quark line which is connected to the photon), the decay constants of the mesons, and the GPD and DAs. 
 
For example, for two longitudinal $\rho^0$ mesons produced from a proton, this prefactor reads
\begin{equation}
f(\{244,1,1,\{3,2\}\},p,p,\{\rho^0,L\},\{\rho^0,L\})=f_{\rho}^2 \,\phi_1(v)\phi_2(z)( {Q_d}   {H_d}(x) +Q_u H_u (x))\,.
\end{equation}
One recognises the DAs $\phi_1(v)$ and $\phi_2(z)$, the square of the rho meson decay constant $f_{\rho}$, the quark electric charges $Q_u$ and $Q_d$, and the vector GPDs $H_u(x)$ and $H_d(x)$ corresponding to the quark flavours $u$ and $d$. The factor $1/2$ that is supposed to appear due to the square root of 2 in the $\rho^0$ wave function is in fact compensated by the equal contribution from the $u$ and $d$ parts of this wave function. Indeed, this diagram belongs to first topology in \FIG\ref{fig:topologies}, where the photon is attached to the lower blob. Thus, the charge seen by the photon, as well as the flavour of the GPD, are completely independent of the flavours taken by the fermionic lines connecting the DAs. Moreover, this specific topology constrains the flavours of both mesons to be the same, both being either $u$ or $d$. In this latter case, the minus sign coming from the two $\rho^0$ wave functions compensate, and therefore, each flavour of the $\rho^0$ meson has the same contribution to this amplitude, including the sign. It is this factor of 2 that compensates the above-mentioned factor of 1/2.

Therefore, if one chooses a specific process, all prefactors are replaced in the list of diagrams. Often, most of them actually end up being zero, since a specific process could be completely incompatible with one or more topologies in \FIG\ref{fig:topologies}. For example, the first topology in \FIG\ref{fig:topologies} is incompatible with a transversity GPD (tensor Dirac structure) due to the presence of an odd number of gamma matrices in the trace.

The next step is to integrate the non-vanishing diagrams over the variables $x$, $v$, and $z$, which have support $x \in [-1,1]$, $z,v \in [0,1]$.

\subsection{Numerical integration}
Unlike the previous specific study of exclusive $\pi^+\rho^0_T$ photoproduction in \cite{ElBeiyad:2010pji}, we perform the integration over all variables numerically. One important advantage of doing this is that one is able to investigate different models for the DAs. First, a partial fraction decomposition with respect to $x$ is performed.  Second, each resulting term is further split using the Sokhotski-Plemelj formula, which results in
\begin{equation}
    \int_{-1}^1dx\frac{g(x,v,z)}{x-r(v,z)\pm i\epsilon}=\int_{-1}^1dx\frac{g(x,v,z)-g(r,v,z)}{x-r(v,z)}+\left[ \ln\left(\frac{1-r(v,z)}{1+r(v,z)}\right)\mp i \pi\right] g(r,v,z)\,,
\end{equation}
where $g(x,v,z)$ denotes the numerator of the term, and $r(v,z)$ denotes the real part of the pole in $x$. Note that the only $x$ dependence in the function $g$ corresponds to that of the relevant GPD. The first term on the right-hand side has to be integrated over $x$, $v$, and $z$, whereas the second term on the right-hand side has to be integrated over only $v$ and $z$. 

One has to consider that, during the partial fraction decomposition, the $i \epsilon$ terms are distributed across both the numerator and denominator of the function $g(x,v,z)$. The $i \epsilon$ dependence in the numerator can be safely eliminated. However, some terms become non-integrable in $v$ or $z$ when $i \epsilon$ is eliminated in the denominator. A careful inspection shows that such terms are always of the form
\begin{equation}
    \frac{z-v}{(z-v)^2+i \epsilon}\;\;\text{or}\;\; \frac{z+v-1}{(z+v-1)^2+i\epsilon}\,,
\end{equation}
that is, the problematic part lies on the diagonals of the $(v,z)$ plane. We stress that the divergences that have appeared in $v$ and $z$ as a result of the partial fraction are \textit{spurious} - Indeed, they were absent before performing the partial fractioning procedure. One convenient way to deal with them is to perform a series of \textit{foldings} of the $(v,z)$ integration region, along the lines $z=v$, $z=1-v$, and finally $z=1/2$. This essentially corresponds to  
\begin{align}
\int_{0}^1 dz\int_{0}^1 dv\, \mathcal{A}(z,v)=\int_0^{\frac{1}{2}}\,dz\,\int_0^z\,dv &\bigg(\mathcal{A}(z,v)+\mathcal{A}(v,z)+\mathcal{A}(1-v,1-z)+\mathcal{A}(1-z,1-v)\nonumber\\&+\mathcal{A}(1-v,z)+\mathcal{A}(z,1-v)+\mathcal{A}(1-z,v)+\mathcal{A}(v,1-z) \bigg)\,,
\end{align}
where $\mathcal{A}$ denotes the term that contains spurious divergences in $v=z$ and/or $v=1-z$. In this way, all the $i \epsilon$ factors can be safely eliminated. For example, if we have a generic term of the form $\frac{h(v,z)(z-v)}{(z-v)^2+i\epsilon}$, then after performing the series of folding, one gets
\begin{align}
    &&\frac{h(v,z)(z-v)}{(z-v)^2+i\epsilon}+\frac{h(z,v)(v-z)}{(v-z)^2+i\epsilon}+(z\rightarrow 1-z\;\, \text{and}\;\, v\rightarrow1-v)\underset{v\rightarrow z}{\sim}
    \frac{(v-z)^2\;\left(\frac{\partial{h}(z,v)}{\partial{v}}-\frac{\partial{h(z,v)}}{\partial{z}}\right)}{(v-z)^2+i\epsilon}+...\,,
\end{align}
 where $i\epsilon$ can be safely taken to 0 before integration on the right-hand side. 
 
\section{Numerical results: Unpolarised cross section}



    \begin{figure}
\includegraphics[width=.49\linewidth]{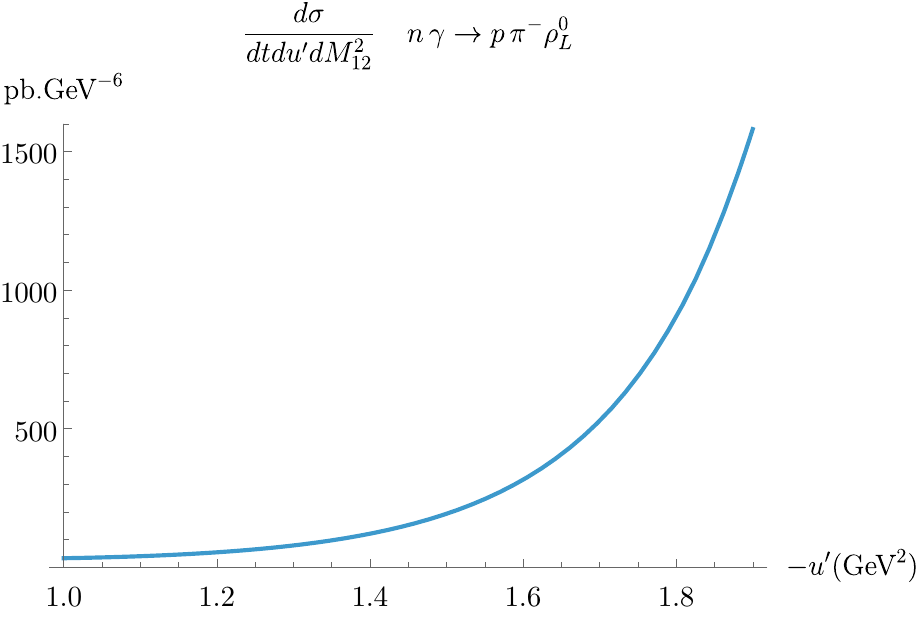}  \includegraphics[width=.49\linewidth]{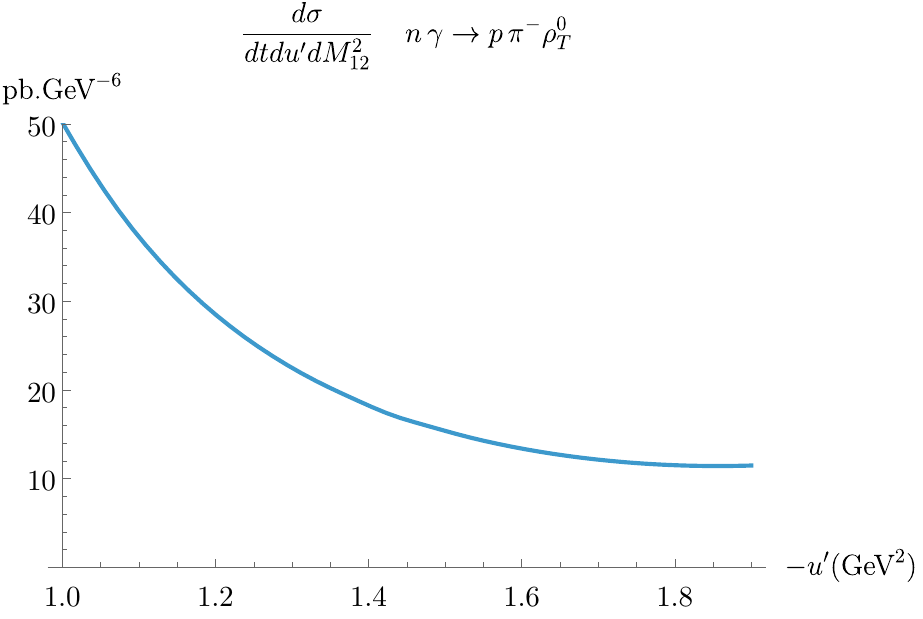}
\\
\includegraphics[width=.49\linewidth]{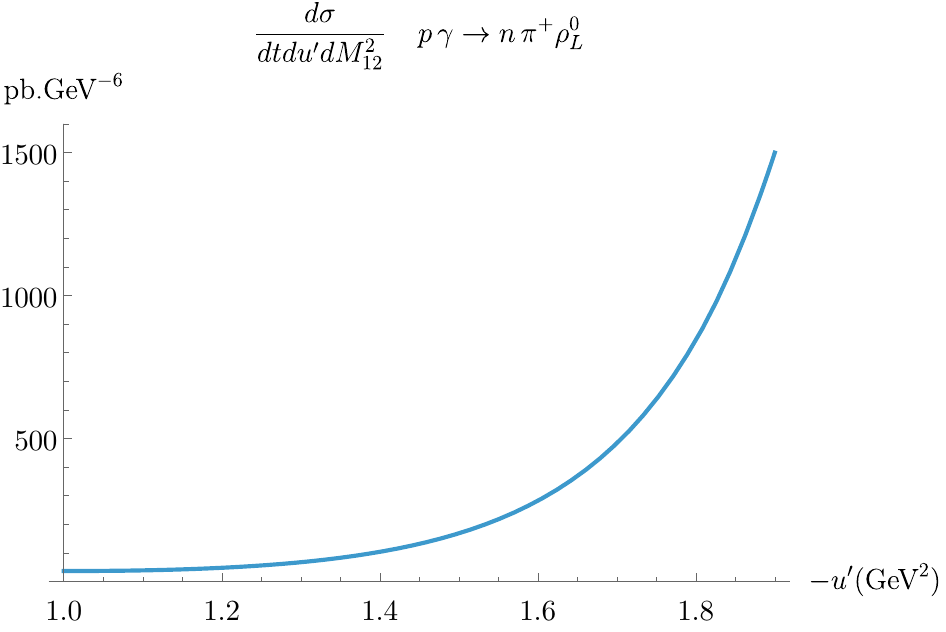}
\includegraphics[width=.49\linewidth]{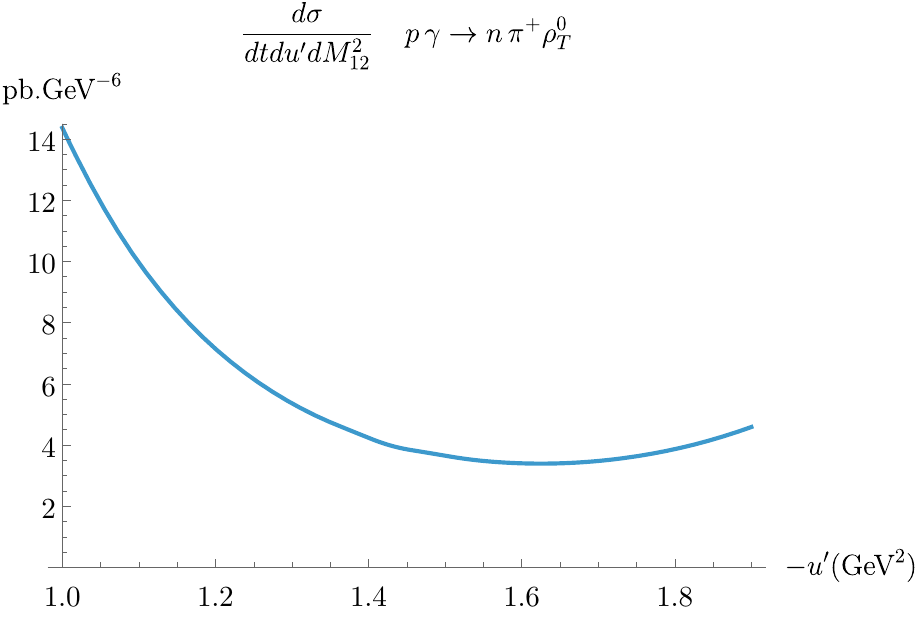}
\caption{Fully differential cross section as a function of $-u'$ for the photoproduction of $\pi^-\rho^0_L$ (top-left), $\pi^-\rho^0_T$ (top-right), $\pi^+\rho^0_L$ (bottom-left) and $\pi^+\rho^0_T$ (bottom-right) at $S_{\gamma N}= \SI{20}{GeV^2}$ and $M^2_{12}=\SI{3}{GeV^2}$.}
\label{fig:differential-cross-section1}
  \end{figure}
    \begin{figure}
         \includegraphics[width=.49\linewidth]{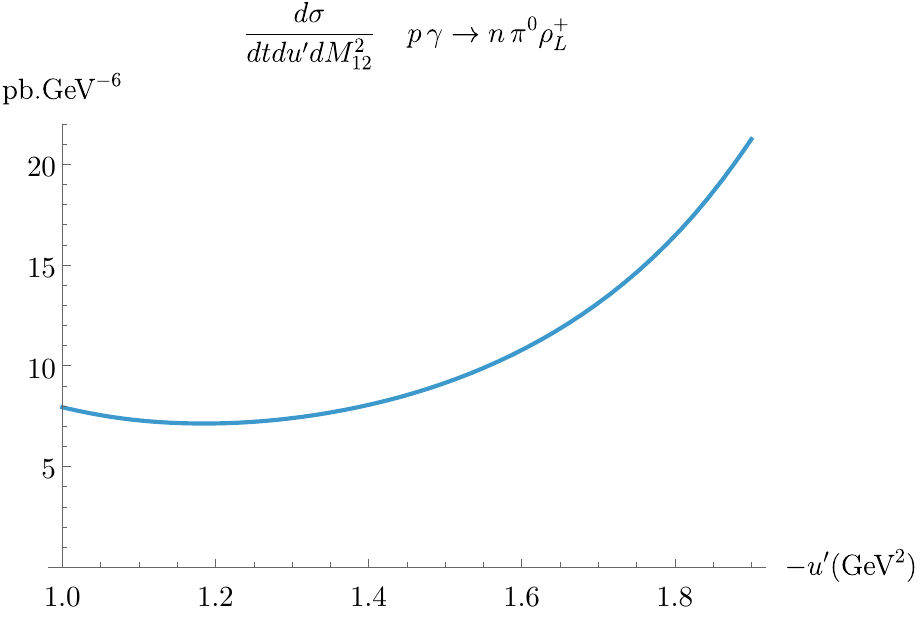}
         \includegraphics[width=.49\linewidth]{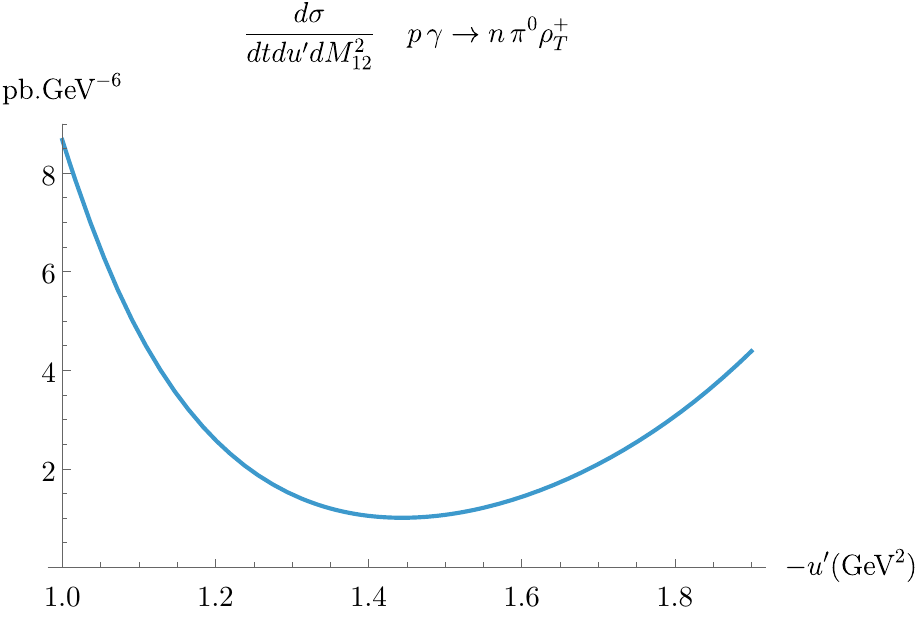} 
         \\
          \includegraphics[width=.49\linewidth]{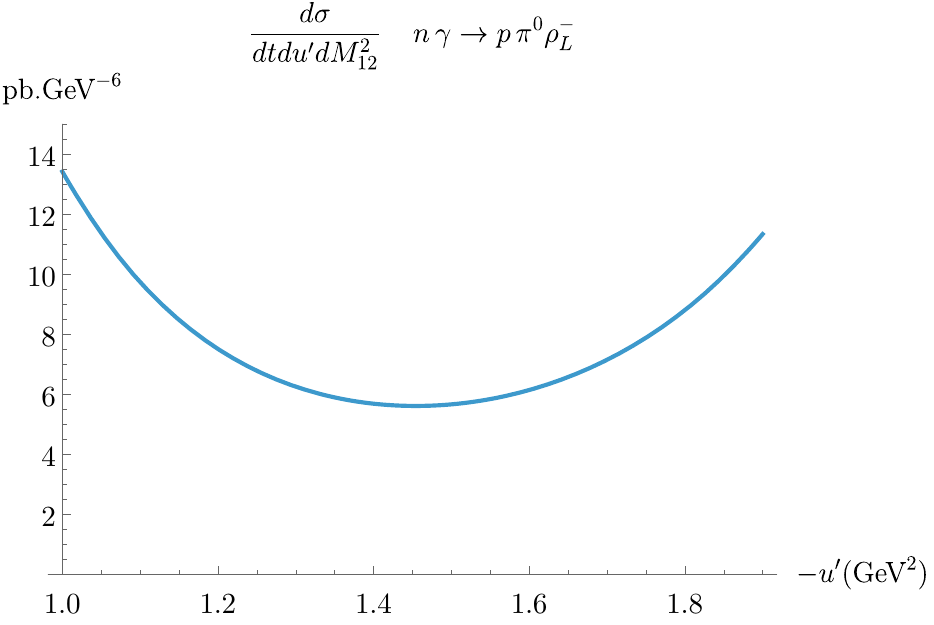} 
         \includegraphics[width=.49\linewidth]{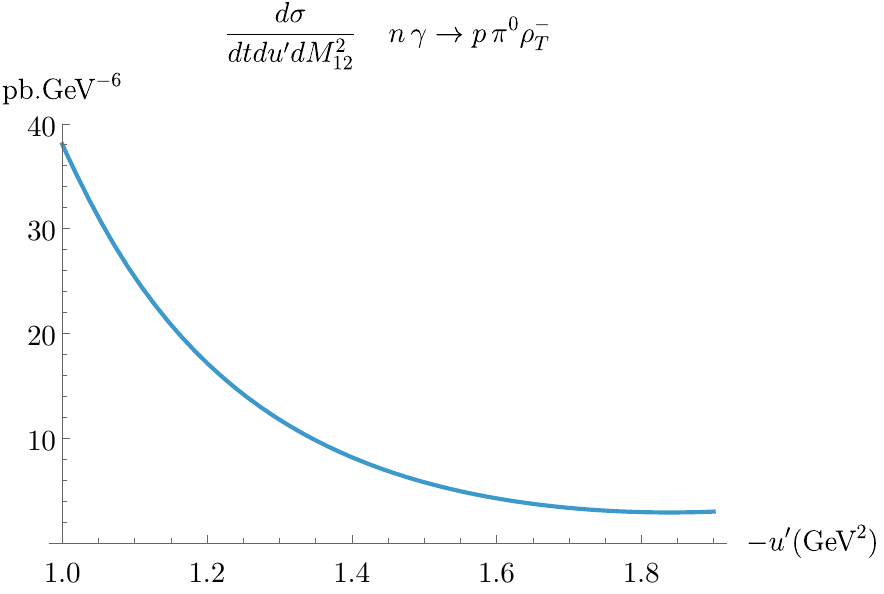} 
         \caption{Fully differential cross sections as a function of $-u'$ for the photoproduction of $\pi^0\rho^+_L$ (top-left), $\pi^0\rho^+_T$ (top-right), $\pi^0\rho^-_L$ (bottom-left) and $\pi^0\rho^-_T$ (bottom-right) at $S_{\gamma N}= \SI{20}{GeV^2}$ and $M^2_{12}=\SI{3}{GeV^2}$.}
\label{fig:differential-cross-section2}
    \end{figure}
        


Let $|\overline{\mathcal{M}}|^2$ denote the amplitude squared averaged over the polarisation of the incoming photon and summed over the polarisations (if any) of the outgoing mesons. The fully differential cross section, expressed in terms of the kinematical variables $-u'$, $-t$ and $M^2_{12}$, is
\begin{equation}
    \frac{d\sigma}{d(-t)d(-u')dM_{12}^2}=\frac{|\overline{\mathcal{M}}|^2}{32(2\pi)^3S_{\gamma N}M_{12}^2}\,.
\end{equation}
In this proceeding, the values of $S_{\gamma N}$ and $M_{12}^2$  are taken to match the set-up of JLab in Hall B, in which the incoming electron beam has an energy of ${12}\,\mathrm{GeV}$. Constraints on the phase space (see e.g. \cite{Boussarie:2016qop}) require $M_{12}^2$ to lie between 2 and $\SI{9}{GeV^2}$, so we take $M_{12}^2=\SI{3}{GeV^2}$ in the following to illustrate the results.

Some plots of the differential cross section as functions of the Mandelstam variable $u'$ are shown in \FIG\ref{fig:differential-cross-section1} and \FIG\ref{fig:differential-cross-section2}. The range of $(-u')$ is fixed by the two constraints $-u'>1 \;\hbox{GeV}^2$ and $-t'>1 \;\hbox{GeV}^2$ to ensure that the kinematics corresponds to the regime where collinear factorisation is applicable \cite{Duplancic:2022ffo}. While the cross sections of the exclusive photoproduction of a photon-meson pair \cite{Boussarie:2016qop,Duplancic:2018bum,Duplancic:2022ffo,Duplancic:2023kwe} did not exceed a few tens of $\hbox{pb}.\hbox{GeV}^{-6}$, here, the values can reach the order of a thousand, for processes with longitudinal $\rho^0_L$ in the final state.

\begin{figure}[ht!]
 \centering
       \includegraphics[width=.49\linewidth]{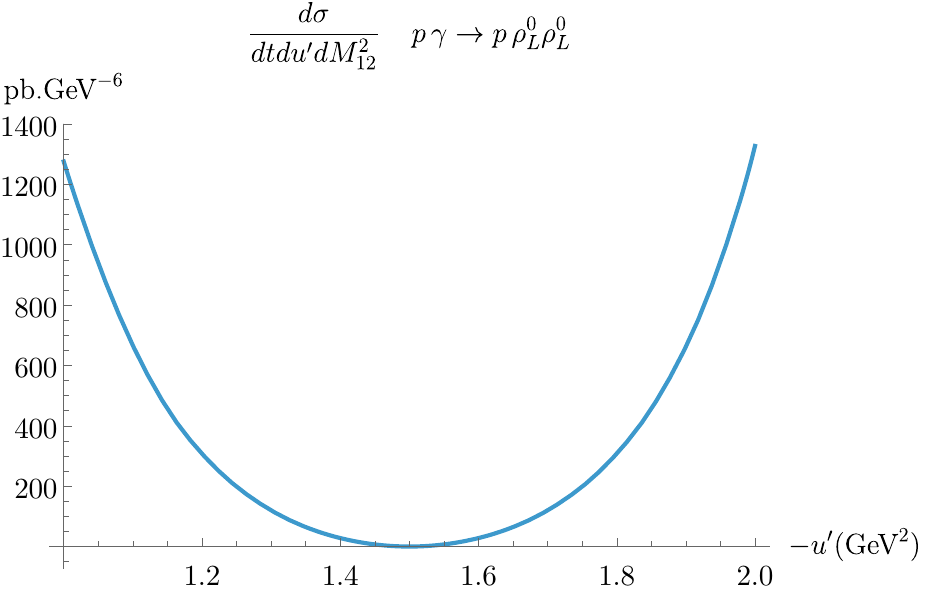} \includegraphics[width=.49\linewidth]{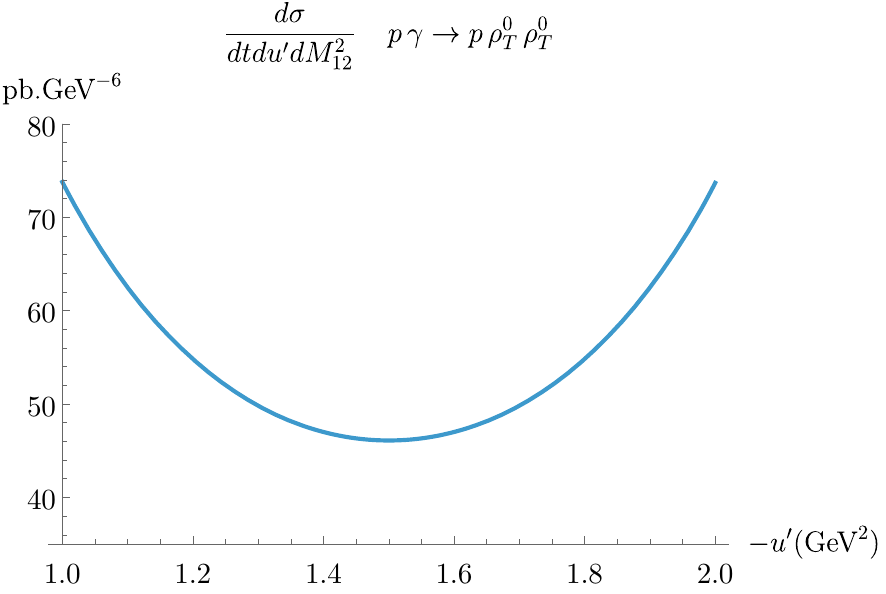}
       \\[6pt]
        \includegraphics[width=.45\linewidth]{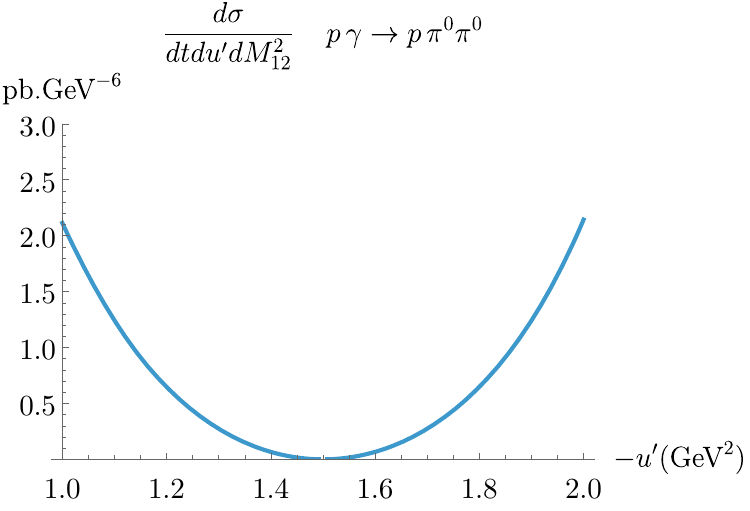} 
          \caption{Fully differential cross sections for the photoproduction of $\rho^0_L\rho^0_L$ (top-left), $\rho^0_T\rho^0_T$ (top-right) and $\pi^0\pi^0$ (bottom) on a proton target at $S_{\gamma N}= \SI{20}{GeV^2}$ and $M^2_{12}=\SI{3}{GeV^2}$. {For the two first processes, the range of $-u'$ has been extended beyond the kinematically allowed range (given the cuts $-u'>1 \;\hbox{GeV}^2$ and $-t'>1 \;\hbox{GeV}^2$, see main text) in order to observe the reflection symmetry of the cross section about the line $-u' = M_{12}^2/2$.}}
    \label{fig:identical-mesons-differential-cross-sections}
\end{figure}

\textit{Meson exchange symmetry} acts as a good check of our results. The cross section obtained for the process $\gamma N\rightarrow N' M_1 M_2$ is related to that of the process $\gamma N\rightarrow N' M_2 M_1$ by the simple transformation of $-u'\rightarrow M_{12}^2-(-u')$, see \EQs\eqref{eq:t-primed} and \eqref{eq:u-primed}.  More precisely, this transformation maps the differential cross section onto its reflection with respect to the line $u'=M_{12}^2/2$. This has been checked for all of our processes. In the special case where $M_1$ and $M_2$ are identical mesons, with the same polarisation, the cross section becomes symmetric with respect to the $-u'=M_{12}^2/2$  axis. This is indeed the case for the processes $\pi^0\pi^0$, $\rho^0_L\rho^0_L$ and $\rho^0_T\rho^0_T$ photoproduction as shown in \FIG\ref{fig:identical-mesons-differential-cross-sections} for a proton target. 
Finally, the same symmetry implies a perfect cancellation of the cross section at $u'=M_{12}^2/2$ for the processes $\pi^0\pi^0$ and $\rho^0_L\rho^0_L$, which is clearly visible from the plots.

\section{Conclusion}
By fully automating the differential cross section computation, we can make predictions for the whole class of $\gamma \, N\rightarrow N' M_1M_2$ processes, where $M_1$ and $M_2$ can be rho mesons or pions of any charge and polarisation, excluding channels which allow the exchange of 2 gluons in the $t$-channel. This paves the way to a broad range of processes for the extraction of both chiral-even  and chiral-odd GPDs. The cross sections that have been computed in JLab kinematics are one or two orders of magnitude above the ones for the  $\gamma \, N\rightarrow N'\gamma M$ processes, which is quite promising. We intend to also compute the linear polarisation asymmetry with respect to the incoming photon and integrated cross sections for JLab, LHC (in ultraperipheral collisions) and future EIC kinematics.

\section*{Acknowledgments}

This work was supported by the GLUODYNAMICS project funded by the ``P2IO LabEx (ANR-10-LABEX-0038)'' and from the "P2I - Graduate School of Physics", in the framework ``Investissements d’Avenir'' (ANR-11-IDEX-0003-01) 
managed by the Agence Nationale de la Recherche (ANR), France. This work was also supported in part by the European Union’s Horizon 2020 research and innovation program under Grant Agreements No. 824093 (Strong2020). This project has also received funding from the French Agence Nationale de la Recherche (ANR) via the grant ANR-20-CE31-0015 (``PrecisOnium'')  and was also partly supported by the French CNRS via the COPIN-IN2P3 bilateral agreement and via the IN2P3 project "QCDFactorisation@NLO". The work of S. N. was partly supported by the Science and
Technology Facilities Council (STFC) under Grant
No. ST/X00077X/1, and 
by the Royal Society through Grant No. URF/R1/201500. L. S. was supported by the Grant No. 2024/53/B/ST2/00968 of the National Science Centre in Poland.

\section*{References}

\bibliography{masterrefs.bib}

\end{document}